\documentstyle[aps,pre,psfig]{revtex}
\begin{document}
\tightenlines
\title{Incommensuration in quantum  antiferromagnetic chain}
\author{Somendra M. Bhattacharjee\cite{eml2}}
\address{Institute of Physics, Bhubaneswar 751 005, India}
\author{Sutapa Mukherji\cite{eml1}}
\address{Universit\"at zu K\"oln, Z\"ulpicher Str. 77, D 50937  
K\"oln, Germany}
\maketitle
\today
\begin{abstract}
  A dimerized quantum Heisenberg or XY antiferromagnetic chain has a
  gap in the spectrum.  We show that a weak incommensurate modulation
  around a dimerized chain produces a zero temperature quantum
  critical point. As the incommensuration wavelength is varied, there
  is a transition to a modulated gapless state.  The critical behaviour
  is in the universality class of the classical
  commensurate-incommensurate (Pokrovsky-Talapov) transition.  An
  analogous metal-insulator transition can also take place for an
  incommensurate chain.
\end{abstract}
\nopagebreak
  
The one dimensional $S=1/2$ quantum antiferromagnetic Heisenberg model
is a prototypical example of quantum fluctuations destroying the
classical ground state\cite{quantum,bethe,klumper,mg,sr,indrani}.
Such a uniform chain, with gapless excitations, can undergo a
spin-Peierls (SP) transition to a lattice dimerized state via the
interplay of spins and lattice distortion, going over to an
incommensurate phase in large magnetic
fields\cite{sp1,sp2,bonner,cugeo3}. The SP state has a gap in the
spectrum.  However, specific heat measurements in the incommensurate
phase of the inorganic insulator CuGeO$_3$\cite{cugeo3}
hint\cite{spheat,bnr} towards a gapless phase.  This, then, raises a
few basic questions: Can a weak incommensuration around a dimerized
Heisenberg (or XY) chain support {\it gapless} excitations, eventhough
the dimerized case {\it does not}?  If it is true, then is there a
quantum critical point (QCP) separating a gapless state and a gappy
state (a state with a gap)?  What would be the nature of the quantum
critical point?

In order to address these issues for the incommensurate phase, and to
gain an understanding of the underlying phenomena, one may
consider a simpler situation of a static case where the coupling
constant follows the incommensurate modulation of the lattice.  We
therefore consider a model hamiltonian
\begin{equation}
  H=\sum_{j=1}^N \left [ \frac{J_{j}}{2}( { S}_{j}^+ {S}_{j+1}^{-}+ 
  {S}_{j}^-{S}_{j+1}^{+}) \right ] + h
  \sum_j{S}_{j}^z ,
\label{hamil2}
\end{equation}
where $S^{\pm}_j=S_{j}^x\pm i S_{j}^y$ are the spin raising and
lowering operators at site $j$, and $h$ is the magnetic field in the
z-direction. The case of interest in this paper is $J_j=J + \delta
\cos (\pi + q)j$, with $\delta \ll J >0$.  This corresponds to a weak
incommensuration around the dimer phase $(q=0)$.  For $\delta =0$, the
system is gapless (uniform chain). For $q=0$, a gap develops in zero
field for $\delta \neq 0$ but a gapless phase is recovered above a
critical magnetic field $h_c$\cite{chitra}.  Our aim is to get the
phases in the $(q,h)$ plane for $\delta \ll 1$, based on low energy
excitations.

We are considering the XY model, because, so far as the gap is
concerned the difference between XY and Heisenberg chain is not
crucial (see Eq. (\ref{freebos})). Admittedly, for the SP problem, the
spin lattice interaction and the magnetic field play important
roles\cite{brazovskii} and correlate\cite{uhrig} $q$ and $h$.  A real
system would follow a particular trajectory $q=q(h)$ in the $(q,h)$
phase diagram. We therefore take $q$ and $h$ as independent
parameters.

Eq. (\ref{hamil2}) also describes, via a Jordon Wigner
transformation\cite{fradkin}, spinless fermions on a one dimensional
lattice - a description we use in this paper.  By defining $
C_j=K(j)S^-_j$ as the fermionic annihilation operator, where
$K(j)=\exp(i\pi\sum_{n=1}^{j-1}S^+_n S^-_n)$ is the nonlocal kink
operator, the spin hamiltonian of Eq. \ref{hamil2} can be written as
\begin{eqnarray}
  \label{jord2}
  H=\frac{1}{2}\sum J_j(C^\dagger_j
  C_{j+1}+C^\dagger_{j+1}C_j)
- h \sum_j C^\dagger_jC_j.
\end{eqnarray}
Note  that the fermions have {\it incommensurate  hopping} rates, and
$h$ acts as a chemical potential. 

A QCP is a phase transition point with a diverging length scale
induced by a change of a parameter of the hamiltonian at zero
temperature.  Since quantum fluctuation is responsible for such
criticality, the dynamic exponent $z$, determining the scaling of time
and space, becomes more important than in thermal critical points
($\tau \sim \xi^z$ with $\tau$ and $\xi$ as characteristic diverging
time and length scales).  In the spin chain context, a wellknown QCP
is the dimerization point $\delta=0$ for $q=h=0$ for Eq.
(\ref{hamil2}) or (\ref{jord2}).  This critical point (separating a
gapless state and a gappy state) corresponds to free fermions with
$z=1$ and a correlation length diverging as $\xi=\delta^{-1}$.  The
spin-spin correlation function $C(r,\delta)=\langle S_i^z
S_{i+r}^z\rangle$ has a scaling behaviour $C(r,\delta) = r^{-\eta}
f(r\delta)$ with $\eta=2$\cite{henelius}.  For the QCP of concern in
this paper, we start from the dimerized gappy state and change the
incommesuration $q$ keeping $\delta$ fixed.

A QCP for $q=0$ at $\delta=0$ ensures the existence of a continuum
limit of the lattice problem.  We use this continuum limit to study
the effect of $q$ on the phase diagram, and in a renormalzation group
framework only the relevant terms need to be kept. To go to the
continuum limit, we adopt the technique of bosonization for the low
energy excitations around the fermi points\cite{born,fradkin,emery}.
Also since wave number is nolonger a good quantum number, we work in
the real space.  The low energy excitations of Eq. (\ref{jord2})
across the fermi surface on the two branches can be described by $L$
(left) or $R$ (right) moving particles with linear spectrum. This
gives $z=1$. In the continuum limit, the Hamiltonian is written in
terms of the (bosonic) phase variables.  The free bosonic theory
corresponding to $J_j=$ constant, $h=0$ turns out to be just the
harmonic lattice hamiltonian, which forms the basis for the subsequent
analysis of the remaining terms such as dimerization, incommensuration
around the dimerized case and the magnetic field. The relevance or
irrelevance of various terms come from their scaling behaviour on a
long length and time scale.  In the renormalization group approach,
the scaling behaviour is obtained by integrating out the short
distance fluctuations in both directions and incorporating their
effects in the parameters of the problem.  This is implemented by
studying the Euclidean version of the problem (i.e., imaginary time,
$t \rightarrow it$).

For low energy excitations, we define new operators $a_j=(i)^{-j} C_j$
to eliminate the fast variation at the Fermi vector. From these, right
and left movers are defined by the chiral transformation
$R(j)=(a_{2j}-a_{2j-1})/\sqrt{2} $ and
$L(j)=(a_{2j}+a_{2j-1})/\sqrt{2}$ so that the continuum version of the
Hamiltonian can be written as
\begin{mathletters}
\begin{eqnarray}
  \label{freecont}
H&=&i \int dx
J [L^\dagger(x)\partial_x L(x)-R^\dagger(x)
\partial_x R(x)] \label{freea}-\\
&&\frac{i}{2}\int dx \ \partial
J(x)(R^\dagger(x)R(x)-L^\dagger(x)L(x))+\label{inta}\\
&&\frac{i}{2}a_0 \int dx \partial^2
J(x)(R^\dagger(x)L(x)-L^\dagger(x)R(x))\label{intb}
\end{eqnarray}
\end{mathletters}
where for simplicity the modulated part of the coupling has been
ignored in Eq. \ref{freea}.  This is justified by the modified Harris
criterion of Ref. \cite{luck} since incommensuration $\delta
\cos(2k_f+q)x$ is bounded.  The notation $\partial J(x)$ and
$\partial^2 J(x)$ denote the appropriate continuum limit of the
discrete difference of the modulation of the couplings.

The bosonic operators are obtained from the phases\cite{sp2,nakano} of
$R(x)$ and $L(x)$,
\begin{eqnarray}
R(x)&=&\frac{1}{\sqrt{2\pi a_0}}  \exp(ik_fx-\phi_1(x)),\\
L(x)&=&\frac{1}{\sqrt{2\pi a_0}}\exp(- ik_fx-\phi_2(x)),
\end{eqnarray}
with $k_f=\pi/2$, so that $\theta(x) = i(\phi_1(x) +\phi_2(x))$, and
the conjugate momentum $p$ are related to the density and current
respectively as $\partial \theta/\partial x = R^{\dagger}R +
L^{\dagger}L$ and $p= R^{\dagger}R - L^{\dagger}L$.  With this choice,
the free part of the bosonic hamiltonian (analogous to the first term,
Eq. \ref{freea}) is the standard harmonic chain hamiltonian
\begin{equation}
  \label{freebos}
  H_0=\frac{1}{2\pi}\int dx\left[\frac{1}{K}\left(
  \frac{d\theta}{dx}\right)^2+ K (\pi p)^2\right] 
\end{equation}
in units of $\hbar =1$ and spin wave velocity = 1.  Here, $K=4$ for
the XY model. For the full isotropic Heisenberg hamiltonian or
interacting fermions in the fermionic language, $K=2$.  The second
term, Eq. \ref{inta}, is analogous to a current type term which is
absent if one goes over to the dimer limit (recovering translational
symmetry) and is expected in the incommensurate case on symmetry
grounds (or lack of it).  In the bosonic variables, the current term
in Eq. \ref{inta} is like $\delta\ q \sin Qx\ p(x)$, where $Q=2 k_f +
q$, and it vanishes in the limit $q \rightarrow 0$.  However because
of the oscillatory coefficient with average zero over a period, we
ignore this term in the present analysis.  The third term (\ref{intb})
shows that a spatially varying $J$ gives rise to a umklapp scattering
process which apparently violates momentum conservation by $2k_f$.
This is a special term needed mainly because of the underlying lattice
in the problem.  Considering only slow variations in the bosonic field
$\theta$, the umklapp term can be written\cite{nakano} as $\approx
\frac{\delta}{4\pi} \int dx Q^2\ \cos(\theta(x)-[2k_f-Q]x)$, by
shifting $\theta\rightarrow \theta-\pi/2$.  The bosonized version of
the incommensurate Hamiltonian is therefore (prime denoting spatial
derivative)
\begin{eqnarray}
  \label{hamboson}
  H&=&
\frac{1}{2\pi}\int dx \ [\frac{1}{K}(\theta^\prime)^2+K(\pi
  p)^2]
-\frac{\delta \pi}{4}
 \int dx \cos(\theta(x)+ q x)
\end{eqnarray}
The dimer limit is restored by taking $q=0$ and in this limit with
$k_f=\pi/2$ the umklapp term agrees with Nakano and
Fukuyama\cite{nakano} who obtained a similar term in the presence of a
constant bond alteration.  The above bosonic hamiltonian has
resemblance with the hamiltonian that appears in theory of
incommensurate crystals\cite{pokrovsky}. In those problems $q$ plays
the same role of incommensuration vector.  Such a hamiltonian also
occurs in the Frenkel-Kontorowa model\cite{frenkel} of a harmonic
chain in an external cosine potential.  This similarity shows that the
solitons expected in such cosine potentials are also important in the
spin chain problem.

To study the relevance or irrelevance of the sine Gordon term, we
implement a renormalization group analysis, well documented in Ref.
\cite{wiegmann,pokrovsky}. The basic steps involve obtaining a
functional integral $\int e^{iS(\theta)} d\theta$ where $S(\theta)$ is
a dimensionless action.  By going over to imaginary time $\tau=it$ the
integral is written as a partition function of a classical
two-dimensional problem.  A further shift of the field variable
$\theta(x,\tau )= \theta(x,\tau)- q x$, gives the classical
hamiltonian\cite{comment} as
\begin{equation}
  \label{action2}
H=\frac{1}{\pi K}\int d\tau dx\left[\frac{1}{2}\dot\theta^2 +
\frac{1}{2}(\theta^\prime - q)^2- 
{{\Delta}}\cos\theta\right],
\end{equation}
where ${ {\Delta}}=\delta \pi^2 K/4$.

The RG procedure\cite{wiegmann} for $q=0$ involves decomposing the
cosine term into fast and slowly varying components of $\theta$ for
${{\Delta}}\ll 1$ and then averaging out the fast varying component
with respect to its Gaussian distribution. Absorbing the contribution
from the averaged fast component in ${{\Delta}}$ we can have the
renormalization of the ${{\Delta}}$ as ${ \Delta}_R={
  \Delta}^{1/(1-\beta^2/8\pi)}$, where $\beta^2= \pi K$.  For
$\beta<{\sqrt{8\pi}}$, the sine Gordon term is relevant and yields a
massive theory.  The Heisenberg or the XY model belongs to this
category and, as expected, any dimerization produces a gap in the
spectrum, with the gap scaling as $\Delta_R$, and therefore scaling as
$\delta^{8/(8-K)}$.  With the incommensurate term, the system can
disregard the potential if $q \theta^\prime$ is comparable to the
cosine potential energy.  This gives the phase transition
point\cite{pokrovsky} as $q_c \sim { \delta}^{4/(8-K)}$.  For the XY
model, the transition to the gapless phase is in the same universality
class as the commensurate-incommensurate transition in
two-dimensions\cite{pokrovsky,smb}. For this universality class, the
relativistic invariance leading to $z=1$ is lost.  The spatial length
scale exponent is $\nu=1/2$ with $z=2$\cite{smb2,schulz}.

The results imply that the gap at the fermi level persists for small
$q$ incommensurations around the dimerized phase.  Isolated states
might well appear in the gap but they do not destroy the gap. The
spins basically follow a dimerized lattice.  For larger
incommensuration, i.e. for $q>q_c$, the spins follow the dimerized
lattice over a finite length scale separated by defects or solitons.
Such (spin $1/2$) solitons do exist for the dimerized
chain\cite{nakano}. Though the solitons cost energy, a many soliton
state for large $q$ would be favorable compared to, say, following the
dimerized lattice over the whole length (a no soliton state) because
of the energy gain through the ``$q$'' term in Eq. \ref{action2}.  In
this phase with a finite density of solitons, translational invariance
is recovered because these solitons are not bound to the underlying
lattice, and one gets back a massless mode. The density $\rho$ of
these solitons is given by $\langle \theta^\prime\rangle$, and this
density can act as the order parameter for the transition.  In the
gappy phase, $\rho$ is zero (no soliton or domain wall), while $\rho$ is
nonzero in the gapless phase.  The dependence of $\rho$ on the
deviation from the critical point is obtained by matching the chemical
potential to the relativistic fermi energy.  This leads to a density
$\sim (q - q_c)^{\beta}$ with $\beta = 1/2$.  In one dimension, a
length scale $l$ can also be defined from $\rho$ as the average
separation of the solitons.  This length $l$ diverges ($l \sim (q -
q_c)^{-\nu}$) with an exponent $\nu = \beta=1/2$ as the critical point
is reached on the gapless side. Because of the high
energy involved in the soliton formation, there will be no critical
divergences in the gappy state\cite{schulz,smb2,smb}.
The gapless phase is not identical to the free fermion phase, mainly
because of the existence of a length scale $l$ for the average domain
size within which the system follows the dimerized lattice.
E.g., the equal time spin spin correlation
function $C(r,q)$ would have an algebraic decay as for free fermions
or the dimerization QCP with $\eta=2$, but for the incommensurate
QCP, there will be an additional oscillatory factor $\cos(r/l)$ 
\cite{schulz,smb2}. Such oscillatory factors would be detectable in
scattering experiments.

For the spin system, the phase transition induced by the
incommensuration $q$ signals the formation of a band at the fermi
surface (in zero field) through the wandering of the solitons which in
one dimension also act like noninteracting fermions\cite{schulz,smb}.
The low lying spectrum therefore becomes gapless.  A simple
dimensional analysis suggests that the width of the band formed around
the fermi level is $w \sim 1/l^2$.  We, therefore, expect the
bandwidth to vanish as $w \sim |q - q_c|^{2\beta}$,i.e., linearly on
the gapless side.  According to the bosonization rules the local
magnetization is proportional to $\langle \theta^\prime\rangle $ so
that the soliton density also determines the magnetization in the
phase.  Therefore the gappy phase will be nonmagnetic but the
gapless phase is magnetic and the magnetization vanishies with the
Pokrovsky-Talapov exponent $\beta = 1/2$.

So far we have considered the zero field case.  The magnetic field
acts as the chemical potential of the fermions, as seen in Eq.
\ref{jord2}, and so long as the fermi surface is in a band, the
equivalent bosonized hamiltonian will be of the form Eq.
\ref{freebos}, with a renormalized $K$ and the spin wave velocity (
taken to be unity).  Since $h$ couples to $\partial \theta/\partial
x$, such a magnetic field for a dimerized chain would have an
equivalent bosonized hamiltonian as Eq. \ref{action2} with $h$
replacing $q$.  The critical behaviour as the fermi surface reaches the
boundary of the gap will therefore be similar to what we have already
studied.  This has explicitly been shown in Ref. \cite{chitra}. If the
band formed by the solitons lies entirely in the original gap, then by
shifting the fermi level the system can go from a gapless to a gappy state
and then again to a gapless phase.

Our analysis though aimed at the spin problem is equally valid for
hopping spinless fermions on a one dimensional incommensurate lattice
i.e. on a lattice with incommensurate hopping rates, as given by Eq.
\ref{jord2}. We can therefore conclude that for a half filled
lattice, there will be an
insulator-metal transition as the incommensuration wavelength is
varied.  One dimensional incommensurate systems can
therefore be classified as metals or insulators based on the
incommensuration. It might be possible to observe such
a metal insulator transition in an incommensurate crystal
(incommensuration around dimerized lattice) by changing say
temperature or other external parameters that control the
incommensuration of the lattice.

As emphasized at the beginning, our analysis is tied to the dimerized
gappy phase, the existence of a continuum limit, and the perturbative
renormalization group (i.e. $\delta \ll 1$). From the similarity with
the Frenkel-Kontorowa model in the continuum limit, Eq.
(\ref{hamboson}).  and the possibility of a gap in the original
Hamiltonian of Eq. \ref{hamil2} whenever the wavevector $Q=\pi+q$ is a
rational fraction of the fermi wavevector, it is expected to have
stable lock-in phases\cite{pokrovsky,frenkel} around certain rational
fractions. Our procedure would then yield similar critical behaviour
around each such lock-in gappy phase. To get the full phase diagram
(and the possibility of a Devil's staircase\cite{pokrovsky,frenkel})
one needs to study the original lattice model as the strength of the
incommensuration $\delta$ is increased.  Also strong incommensuration
may completely destroy the band structure yielding point spectrum as
known for quasiperiodic systems\cite{pandit}. These remain to be
studied.

To summarize, we have studied the effect of incommensuration in a
quantum antiferromagnetic Heisenberg or XY chain and found a quantum
critical behaviour induced by  incommensurability $q$ around
dimerization. Using renormalization group results we find that, in
zero magnetic field, with the increase of the lattice
incommensurability the massive theory reduces to a massless theory
through a {\it continuous} transition at zero temperature. This
indicates a transition from a nonmagnetic gappy (as in the
dimerized case) to a gapless magnetic phase. The critical behaviour of
this transition is found to be in the same universality class as the
two dimensional commensurate-incommensurate transition, with all
singular features appearing {\it only} on the incommensurate gapless
state.  Correlation functions will have characteristic oscillatory
factors that distinguish the incommensurate phase from the gapless
free fermion phase. This prediction could be tested in synthetic one
dimensional magnetic chains.  We have also pointed out that our
results are equally valid for spinless fermions on special types of
incommensurate crystals.  Therefore the prediction and critical nature
of the metal-insulator transition induced by incommensuration around a
dimerized lattice can be tested in properly fabricated incommensurate
heterojunctions\cite{merlin}.

We thank T. Nattermann for suggesting this problem. SM thanks SFB341
for support. SMB thanks T. Nattermann for hospitality.


\begin{references}
\bibitem{comment} Note that the
current term would have changed the $\dot\theta$ term as well, and the
general form of the  square gradient term would have been $ (\nabla
\theta - {\bf A}(x))^2$ with ${\bf A}$ depending only on the space  
coordinate.
\end{references}
\end{document}